\documentclass[preprints,article,accept,moreauthors]{Definitions/mdpi} 

\firstpage{1} 
\pubvolume{1}
\issuenum{1}
\articlenumber{0}
\pubyear{2026}
\copyrightyear{2026}
\datereceived{ } 
\daterevised{ } 
\dateaccepted{ } 
\datepublished{ } 

\newtheorem{condth}{Condition}

\newenvironment{condition}[2][]{\begin{condth}[#1]\label{#2}}{\end{condth}}

\def\>{\rangle}
\def\<{\langle}

\DeclareMathOperator{\erf}{erf}
\DeclareMathOperator{\tr}{tr}

\Title{Reverse Physics for Quantum Mechanics}

\Author{Gabriele Carcassi $^{*\orcidA{}}$, Tobias Thrien $^{\orcidB{}}$ and Christine A. Aidala $^{\orcidC{}}$}

\AuthorNames{Gabriele Carcassi, Tobias Thrien and Christine A. Aidala}

\address[1]{%
Department of Physics, University of Michigan, Ann Arbor, Michigan 48109, USA}

\corres{Correspondence: carcassi@umich.edu}

\abstract{Reverse Physics is a methodology that breaks physical theories into separate mathematical and physical conditions to establish their logical relationships. To showcase the power of the methodology, we present several results for quantum mechanics and their related insights. The standard Hilbert-space formulation conflicts with basic physical requirements, while a minimal topological modification can solve these problems. The ensemble space, rather than the pure-state space, distinguishes classical from quantum systems. The Born rule is an additional assumption linking orthogonality, mutual exclusivity and information entropy. Under explicit background conditions, unitary evolution is equivalent to deterministic and reversible evolution. Nonselective projective measurements can be characterized as Lindblad equilibration processes, while unitary evolution can be characterized as a limit of infinitesimal projective processes. Classical mechanics is recovered as the high-entropy limit of quantum mechanics, and every quantum state, pure or mixed, is a dynamic, spectral and thermodynamic equilibrium. These results are self-contained, use the standard vector-space representation and can thus be used as common tools and constraints for teaching, interpretations, reconstructions and future theories.}


\begin{document}

\section{Introduction}

The core question of this special issue is, ``What kinds of entities does quantum mechanics refer to?'' The fact that, one hundred years after the birth of the theory, this question still has no conclusive answer \cite{SchlosshauerKoflerZeilinger2013,JedlickaEtAl2025} should force us to step back and ask, ``What methodology is being used to answer this question? Is it appropriate? How does it compare to other methodologies used in foundational disciplines?''

Interpretations and reconstructions of quantum mechanics seem to be the most used methods \cite{Myrvold2022,Grinbaum2007,Berghofer2024}, but they both suffer similar issues. First of all, they proceed as though we have already identified the appropriate mathematical entities that \emph{can} map one-to-one to physical entities, and it is just a matter of giving them physical meaning or an alternative definition. But known open problems show that we are not in that position \cite{CarcassiCalderonAidala2025,JaffeWitten2006,Gallavotti2006,Heathcote1990,Ruetsche2011,EarmanFraser2006,Redei1996,Toader2021}. In fact, we argue that a better understanding of the physics is required to inform what the correct mathematical structures should be. A second problem is that, most of the time, the ontology of an interpretation or the different starting points of a reconstruction are chosen, effectively, as a matter of preference \cite{SchlosshauerKoflerZeilinger2013,JedlickaEtAl2025,Grinbaum2007,Berghofer2024}. One may privilege information-theoretic aspects; another wants a realist answer to the measurement problem; yet another focuses on mathematical rigor \cite{Berghofer2024,AlloriEtAl2008,KronzLupher2024}, with the physics, in some cases, only appearing after 120 pages of theorems. We do not contest that everyone is entitled to their approach. The issue is that, in practice, even understanding exactly the results and limitations of one approach requires significant investment, and it is even more difficult to make comparisons. A third problem is that many claims remain specific to the approach and cannot serve as a common basis. For example, if one introduces entities with features that, by their definition, are not, directly or indirectly, experimentally accessible, their relevance to experimental physics will be limited and not universal. On the technical side, some moves work very well in finite-dimensional systems, but do not carry over to infinite-dimensional ones. Some work in the opposite direction \cite{Ruetsche2011}. Interpretations and reconstructions that are based on those moves cannot provide insights that are general. Our conclusion, then, is that these methodologies are unlikely to provide incontrovertible results that can reach a broad consensus.

If we look at the foundations of mathematics, while there are very different schools of thought on what should be accepted as mathematics, there is tacit agreement that there must be some logical system that provides well-defined sentences and rules of inference, and a set of axioms on the objects those sentences describe \cite{Horsten2022,Ferreiros2008}. One may accept or not accept the law of the excluded middle or the Axiom of Countable Choice, but all will agree that, if one or the other is accepted, it will have consequences. The aim of Reverse Physics \cite{aop-phys-ReversePhysics} is to perform a similar analysis for all physical theories, providing stable common results that one can use freely in their interpretations, reconstructions or as insights for future theories. With that in mind, Reverse Physics aims to be practical. Each result should be expressible and provable with standard tools (e.g. vector space formulation, phase-space formulation), accessible to anyone with an advanced undergraduate background in math and physics, and cleanly extractable from the overall framework.

For the past five years, we have been working on refining the methodology, to make it as close as possible to the practice of mathematical logic while retaining the ability to work with informal objects, as required by physics. In this paper, we will present the status of this process, through a series of results that have functioned as different test cases. Most of the results presented are new, and the others were reformulated using the refined methodology. We will provide only proof sketches, leaving the  full details in technical briefs available through our website \cite{AssumptionsOfPhysicsWebsite} and other works. The goal, in fact, is to showcase the methodology to those that are both interested in the foundations of physics and well-versed in formal and informal logic, making philosophers of physics particularly suited to its review and criticism.

In section two we will give a brief introduction to Reverse Physics. In section three we will see that Hilbert spaces fail to respect basic physical conditions. In section four we see a way to partly fix the vector space formulation. In section five we see that it is the convex structure of statistical ensembles that differentiates quantum from classical mechanics. In section six we see that the Born rule is independent of said structure, and is equivalent to assuming orthogonality and mutual exclusivity coincide. In section seven we see that unitary evolution is exactly deterministic and reversible evolution. In section eight we see that nonselective projective measurements are equilibrations, that they can be understood as the limit of an open evolution and that unitary evolution can be seen as an infinite sequence of infinitesimal measurements. In section nine we see that classical mechanics can be recovered as the high-entropy limit of quantum mechanics. In section ten we see that all states, including pure ones, are dynamic, statistical and approximately thermodynamic equilibria. The results follow from standard textbook mathematics and physics, without interpretation-specific assumptions, thus clarifying what we are committing ourselves to when we use the theory. As such, they can inform all interpretations and reconstructions.

\section{Introduction to Reverse Physics}

In the foundations of mathematics, a common fruitful strategy is to analyze the logical dependencies of different statements over a set of preconditions \cite{Simpson2009,Stillwell2018}. A famous example is the following \cite{Pinter2014,Schechter1997}:
\begin{Theorem}
	Over Zermelo-Fraenkel set theory, the following are equivalent:
	\begin{description}
		\item[AoC] (Axiom of Choice) For every non-empty set $X$ there exists a function $f$ that assigns to each nonempty set $S \subseteq X$ some
		representative element $f ( S ) \in S$;
		\item [WOP] (Well-Ordering Principle) Every set can be well-ordered;
		\item [ZL] (Zorn's Lemma) Every non-empty partially ordered set $P$ with the property that every chain in $P$ is bounded has a maximal element.
	\end{description}
\end{Theorem}
These types of investigations allow one to better understand the structure of mathematics and clarify what each starting point commits us to, particularly when intuition can be faulty. As Jerry Bona declared \cite[p. 145]{Schechter1997} regarding intuition about these three equivalent statements:
\begin{quote}
	``The Axiom of Choice is obviously true, the well-ordering principle obviously false, and who can tell about Zorn's lemma?''
\end{quote}
In Reverse Physics, we similarly want to break physical theories into a set of mathematical and physical conditions and study their logical relationships.

For example, if we set
\begin{condition}[Classical states for 1 DOF]{CST-1D}
	The state of the system is represented by position and momentum $(q,p) \in \mathbb{R}^2$. The count of states is given by the Liouville measure $\mu(U) = \int_U dq dp$. The ensemble space is represented by the set of probability distributions $\rho(q, p)$ (i.e. probability measures that are absolutely continuous with respect to the Liouville measure).
\end{condition}
then we have \cite{aop-book,aop-phil-Hamiltonianinformation}
\begin{Theorem}
	Over \ref{CST-1D}, the following are equivalent:
	\begin{description}
		\item[HM-1D] (Hamiltonian mechanics) The equations of motion are given by $d_t q = \partial_p H$ and $d_t p = - \partial_q H$.
		\item [DI-SYMP] (Symplectomorphism) The evolution leaves the symplectic form $\omega= dq \wedge dp$ invariant;
		\item [DI-POI] (Poisson bracket invariance) The evolution leaves the Poisson bracket $\{f,g\} = \partial_q f \partial_p g - \partial_p f \partial_q g$ invariant;
		\item [DR-DIV] (Incompressible flow) The displacement field $S^a = (d_t q, d_t p)$ is divergenceless: $\partial_a S^a = 0$;
		\item[DR-DEN] (Volume conservation) The Liouville measure is invariant;
		\item[DR-EV] (Determinism and reversibility) Past and future states are mapped one to one (i.e. the count of states is preserved);
		\item[DR-THER] (Thermodynamic reversibility) The evolution is thermodynamically reversible (i.e. the entropy given by $\log \mu(U)$ is conserved);
		\item[DR-INFO] (Conservation of information) The information about the system is conserved (i.e. the information entropy given by $- \int \rho \log \rho dq dp$ is conserved).
	\end{description}
\end{Theorem}
The first five conditions are purely mathematical. The last four are more physical. The bridge between them is possible because condition \ref{CST-1D} clearly states the connection between the math and the physics. This allows the proofs to bridge mathematical and physical arguments. All conditions that have the same prefix represent the same set of physical assumptions. For example, all DR conditions are different specifications of the same assumption of determinism and reversibility, linking different notions of vector calculus, differential geometry, statistics, measure theory, thermodynamics and information theory. On the other hand, DI is stronger as it also includes assumption IND of independence of all degrees of freedom. In other words, DI = DR + IND. If we only have a single DOF, then IND is automatically true as a single DOF is a trivial set of independent DOFs, therefore DI and DR are equivalent. If we have multiple DOFs, however, DI is a stronger condition. Therefore, the conditions are equivalent over \ref{CST-1D}, but not over CST-ND.

These are the types of systematic investigations that we want to extend to all physical theories. We want to be able to establish logical relationships, but these may depend on the background theory chosen. Some conditions, in fact, may be applicable across theories. For example, DR-EV and DR-INFO will reappear later when discussing unitary evolution. The naming convention, then, helps us keep track conceptual relationships that span across different theories, finding those that survive across all physical theories.

As we will see, Reverse Physics for quantum mechanics is significantly harder: the standard postulates do not neatly correspond to separate physical assumptions; in fact, there is a significant mismatch between the mathematical definitions and the physical objects they supposedly represent. No perfect interpretation or reconstruction of quantum mechanics, then, is possible within the current established mathematical framework, that maps, and only maps, valid physical objects to valid mathematical ones. Since this final goal is not readily obtainable, we first need to disentangle some of the key pieces of quantum theory and get a sense of what the moving parts are.\footnote{Reverse Physics is effectively applying notions of reverse engineering to physical theories.} We designed Reverse Physics to allow one to carve out precise results within a larger imperfect context, so that one can use those conclusions as the bedrock for further analysis.

The examples that follow illustrate the range of logical relationships that Reverse Physics can expose: consistency or inconsistency with physical requirements, implication, independence and equivalence between mathematical and physical conditions.

\section{The problem with Hilbert spaces}

\textbf{Unphysicality of Hilbert spaces.} A successful interpretation or reconstruction of quantum mechanics relies on the suitability of the mathematical structure to represent the corresponding physical objects. However, the following result shows that the Hilbert space formulation of quantum mechanics presents problems in that respect \cite{CarcassiCalderonAidala2025}.

\begin{condition}[Continuity of measurable quantities]{[B]COBS}
 	Small state changes lead to small changes of measurable quantities. That is, the expectation of a quantity is a continuous function of the state.
\end{condition}
 
\begin{condition}[Measurable quantities well-defined in all frames]{[B]WDQ}
 	All states can equivalently be expressed in all frames. That is, if a quantity is well-defined in one frame, the corresponding quantity in another frame will be well-defined.
\end{condition}

\begin{condition}[Distinguishability of different systems]{[B]STDIST}
 	Different physical systems have a different mathematical representation. That is, the mathematical representation of different physical systems must have enough structure to tell them apart.
\end{condition}

Note that the conditions are marked with [B]. In Reverse Physics, this marks \textbf{base conditions}, conditions that must hold in any well-formed physical theory. For example, if \ref{[B]COBS} does not hold, refining the measurement of an observable does not help narrow down the physical state. If \ref{[B]WDQ} does not hold, say position is valid in one frame but not another, then the physical description is not frame independent, violating the principle of relativity. Lastly, if \ref{[B]STDIST} does not hold, the mathematical representation does not fully capture the physics at hand.  The collection of all base conditions forms the base theory for all of physics, and every physical theory is a specialization of this base theory. One of the long-term goals of Reverse Physics is to find such a base theory.

Characterizing the use of Hilbert spaces in quantum mechanics leads to the following
\begin{condition}[Quantum states and observables]{QSTO:H}
	The state space of a quantum system is represented by the projective space $\mathcal{P}(\mathcal{H})$ of a separable complex Hilbert space $\mathcal{H}$. The ensemble space is represented by the density operators $\mathcal{D}(\mathcal{H})$. Observables are represented by self-adjoint operators. The conditional probability is given by the Born rule $p(\psi|\phi) = \frac{\< \phi | \psi \>\< \psi | \phi \>}{\< \psi | \psi \>\< \phi | \phi \>}$. The entropy is given by the von Neumann entropy $S(\rho) = -\tr(\rho \log \rho)$.
\end{condition}
\begin{Theorem}[Unphysicality of Hilbert spaces in quantum mechanics]
	Condition \ref{QSTO:H} is inconsistent with \ref{[B]COBS}, \ref{[B]WDQ} and \ref{[B]STDIST}.
\end{Theorem}

Since \ref{QSTO:H} applies to all quantum systems and base conditions apply to all physical theories, a single counterexample suffices to prove the inconsistency. For a violation of \ref{[B]COBS}, suppose $|i\>$ represents the $i$th energy level of a harmonic oscillator. The sequence $\sqrt{\frac{j-1}{j}}|0\> + \sqrt{\frac{1}{j}}|j\>$ converges to the ground state as $j \to \infty$, while the expectation of $N$ converges to $1$. \emph{Convergence to a state does not imply that observable expectations converge to those of that state.} This is precisely because, in the topology of a Hilbert space, only bounded operators are continuous and therefore all unbounded operators are discontinuous.

For a violation of \ref{[B]WDQ}, suppose a one-dimensional space, for simplicity, with two observers $X$ and $Y$ linked by the coordinate transformation $y(x) = \tan \left(\frac{\pi}{2} \erf(x) \right)$. A Gaussian wave packet $\psi(x) = \sqrt{\frac{e^{-x^2}}{\sqrt{\pi}}}$ in the first frame will become $\phi(y) = \sqrt{\frac{1}{\pi (y^2 + 1)}}$ in the second frame. The wavefunction will be in the domain of the position $X$ of the first observer but not in the domain of the position $Y$ of the second observer, since $\<Y\phi(y) \mid Y\phi(y)\>$ does not converge. \emph{Position is not defined in all frames.} The cause is that, in a Hilbert space, unbounded operators are not defined on the whole space, and their domains can, in principle, even be disjoint.

For a violation of \ref{[B]STDIST}, consider the space $L^2(\mathbb{R})$ for a single DOF and the space $L^2(\mathbb{R}^n)$ for $n$ DOF. Since they are both infinite-dimensional and separable, they are isomorphic as Hilbert spaces. Therefore, mathematically, the spaces are indistinguishable. \emph{The state space of a single particle is indistinguishable from the space of all the particles in the universe.} The cause is that the definition does not capture enough information to characterize the physical system.

\textbf{Accidents of time.} The use of Hilbert space in quantum mechanics seems to be dogmatically accepted by the majority of physicists, mathematicians and philosophers of physics, even those that work on foundational issues \cite{Halvorson2011}, even though it is known that there are problems, especially in quantum field theory, and even though von Neumann, the creator of the formalism, later stated \cite{Redei1996}:
\begin{quote}
	``I would like to make a confession which may seem immoral: I do not believe absolutely in Hilbert space any more.''
\end{quote}

Many consider these problems, such as the lack of convergence above, as merely mathematical ones. But as Aliprantis and Border state \cite[p. 163]{AliprantisBorder2006}:
\begin{quote}
	``Since there is more than one topology of interest on an infinite dimensional space, the choice of topology is a key modeling decision that can have economic as well as technical consequences.''
\end{quote}
It is fascinating that this point is better understood within economics than within physics.

\textbf{How to proceed.} The conclusion is that it is not currently possible to provide a complete interpretation or a reconstruction of quantum theory. The mathematical objects cannot faithfully correspond to physical entities because they do not possess the right properties. The complication is that finding an appropriate mathematical formalism, understanding the physics being described and having a clear conceptual picture are not independent problems. Therefore, we have to abandon the idea, at least for now, that we will solve all problems. But we do need a methodology that allows us to find, characterize and solve individual problems in a way that will fit into the complete solution. This, in our opinion, is the key methodological problem in the foundations of physics, which Reverse Physics aims to address.

\section{Topology and experimental verifiability}

\textbf{A more physical definition.} While Hilbert spaces clearly have issues, it should be clear that the formalism is very successful anyway. Simply discarding it, then, would be utterly foolish. Here is a minimal modification that addresses the issues presented above.
\begin{Definition}
	A \textbf{quantum vector space} is a complete second-countable complex topological vector space $V$ that satisfies the following:
	\begin{itemize}
		\item $V$ is equipped with an inner product $\< \cdot , \cdot \> : V \times V \to \mathbb{C}$
		\item $V$ is equipped with a set of defining observables $\{X_i\}_{i \in I}$, which are everywhere-defined self-adjoint linear operators $X_i : V \to V$, satisfying a given Lie algebra of commutation relationships $\frac{[X_i, X_j]}{\imath \hbar} = \sum\limits_k C_{ij}^k X_k$
		\item the topology is the coarsest vector space topology making the defining observables and the inner product continuous.
	\end{itemize}
\end{Definition}
\begin{condition}[Quantum states and observables]{QSTO:V}
	The state space of a quantum system is represented by the projective space $\mathcal{P}(V)$ of a quantum vector space $V$. Observables are represented by continuous everywhere-defined self-adjoint operators. The conditional probability is given by the Born rule $p(\psi|\phi) = \frac{\< \phi | \psi \>\< \psi | \phi \>}{\< \psi | \psi \>\< \phi | \phi \>}$.  The entropy is given by the von Neumann entropy $S(\rho) = -\tr(\rho \log \rho)$.
\end{condition}
\begin{Theorem}[Partial physicality of quantum vector spaces]
	Condition \ref{QSTO:V} implies \ref{[B]COBS}, \ref{[B]WDQ} and \ref{[B]STDIST}.
\end{Theorem}

Note that conditions \ref{QSTO:H} and \ref{QSTO:V} have a colon. In Reverse Physics, this indicates alternative versions of the same condition. In fact, note that \ref{QSTO:V} still postulates a complex vector space, with a topology and an inner product that is continuous in that topology. The changes are, by design, as little as possible.

We can see that \ref{QSTO:V} guarantees the previous base conditions. Observables and the inner product must be continuous, which addresses \ref{[B]COBS}. Observables must be defined for all states; therefore, it is not possible for two frames to use different observables with different domains, which addresses \ref{[B]WDQ}. Moreover, two different systems will have either a different set of defining observables or different commutation relationships, which solves \ref{[B]STDIST}.

\textbf{The physical meaning of topology.} The fix is not merely technical; it is also conceptual. Topologies in physical theories capture experimental verifiability, with open sets corresponding to experimentally verifiable statements \cite{aop-book,aop-math-topologydistinguishability,Kelly1996,GeninKelly2017,Vickers1989}. Therefore, if a system is fully characterized by a set of defining observables and their probability distributions, it makes sense that its topology (i.e. how we identify states experimentally) is fully characterized by the observables and the Born rule (i.e. the inner product). The above definition, then, is both justified on physical grounds and works in practice.

Now, the above is only a partial solution that solves some problems and opens others.\footnote{For example, we need a way to define density operators without direct reference to the spectral theorem. Also, dimensional analysis is, strictly speaking, incompatible with observables being linear operators.} But the result stands, and it starts giving us some insight into why the problem exists and how, in a final solution, it should be solved. The topology must be connected to observables, and the definition of the state must include the Lie algebra of the observables. For the rest of the paper, we will go back to analyzing the Hilbert space formalism itself, identifying what parts are actually physically interesting and significant.

\section{Pure states are not enough}

\textbf{States without inner product.} Since a Hilbert space is a fairly sophisticated mathematical object, we want to peel back the mathematical definition, layer by layer, to understand what mathematical requirements represent what physical conditions. The first thing we should note is that the construction of the projective space neither depends on nor requires the inner product. Is stating that states are rays of a complex vector space enough to exclude classical mechanics? It turns out it is not.\footnote{A consequence of states being represented by rays is that the linearity of the underlying vector space must be understood merely as a mathematical convenience. Non-linear maps that preserve rays exist, and therefore they would preserve all the physics. A non-linear representation, then, is possible but uselessly inconvenient \cite{Carcassi2026Nonlinear}.}

\begin{condition}[Quantum state space only]{QP-RAY}
	The state space of a quantum system is represented by the projective space $\mathcal{P}(V)$ of a complex topological vector space $V$.
\end{condition}

\begin{condition}[Classical state space only]{CP}
	The state space of a classical system is represented by a symplectic manifold $(X,\omega)$.
\end{condition}

\begin{condition}[Classical-Quantum exclusivity]{CQE}
	A system cannot be both classical and quantum.
\end{condition}

\begin{Theorem}
	Condition \ref{QP-RAY}+\ref{CP} is consistent with $\neg$\ref{CQE}.
\end{Theorem}

To prove the theorem, it suffices to show that there is a system, classical or quantum, whose state space can be represented both by a projective space and by a symplectic manifold. Finite-dimensional complex projective spaces are symplectic manifolds; therefore, any finite-dimensional quantum system will work \cite{AshtekarSchilling1999}. Furthermore, the state space of a classical magnetic dipole is a sphere, which can be represented by the complex projective line (i.e. a two-state quantum system) \cite{David2015,AshtekarSchilling1999}. In general, the space of pure states is not enough even to distinguish between classical and quantum states.

\textbf{Adding ensembles.} While the state spaces for a magnetic dipole and a qubit are the same (i.e. a two-dimensional sphere), the ensemble spaces are different. In the classical case, the space of ensembles is represented by all possible probability densities over the sphere. In the quantum case, the space of ensembles corresponds to the interior points of the ball \cite{BengtssonZyczkowski2017}. In other words, the spaces of ensembles are enough to differentiate classical and quantum systems.

\begin{condition}[Quantum ensemble space]{QE-DENS}
	The state space of a quantum system is represented by the projective space $\mathcal{P}(\mathcal{H})$ of a separable complex Hilbert space $\mathcal{H}$. The ensemble space is represented by the density operators $\mathcal{D}(\mathcal{H})$.
\end{condition}

\begin{condition}[Classical ensemble space]{CE}
	The state space of a classical system is represented by a symplectic manifold $(X,\omega)$. The ensemble space is represented by the set of probability distributions $\rho(q^i,p_j)$ (i.e. Radon--Nikodym derivatives of probability measures that are absolutely continuous with respect to the Liouville measure).
\end{condition}

\begin{Theorem}
	Condition \ref{QE-DENS}+\ref{CE} implies \ref{CQE} for every non-trivial system (i.e. the state space has more than one state).
\end{Theorem}

Classical ensembles are characterized by a single distribution over states while non-trivial quantum ensembles admit infinitely many, equivalent decompositions \cite{BengtssonZyczkowski2017,Kupsch1998}. If the system is non-trivial, we can find a non-trivial ensemble and check for multiple decompositions to determine whether the system is classical or not.

\textbf{Convex space of ensembles.} The space of ensembles, in fact, must have a \textbf{convex structure} where convex combinations (i.e. $\sum_i p_i \rho_i$ with $\sum_i p_i = 1$ and $p_i\geq 0$) represent statistical mixtures. In this structure, we can see that quantum ensembles allow multiple decompositions in terms of pure states,\footnote{Another result in Reverse Physics is that superpositions of state vectors exactly correspond to multiple decompositions of ensembles \cite{CarcassiCalderonAidala2025,Carcassi2026Superpositions}.} while classical ensembles do not.\footnote{The space of classical ensembles is akin to a Choquet simplex, while the quantum one is not.} This last property is what allows us to treat a classical system as being in a particular well-defined, if unknown, state; the lack of this property is exactly why this does not work in quantum mechanics \cite{Kupsch1998}. The uncertainty over two particular states becomes uncertainty over any superposition of those two states.

\section{Orthogonality and mutual exclusivity}

\textbf{Inner product and orthogonality.} Note that \ref{QE-DENS} implicitly requires the inner product through the definition of trace-class operators. Does this implicitly require the Born rule as well? The following result is useful for that analysis \cite{CarcassiCalderonAidala2025,Carcassi2026BornEntropy}.
\begin{condition}[Probabilities of statistical mixtures]{[B]PMIX}
	The conditional probability of a statistical mixture is the statistical mixture of the conditional probabilities. That is,
	$p(\psi\mid\sum_i \lambda_i\rho_i)=\sum_i \lambda_i p(\psi\mid\rho_i)$.
\end{condition}
\begin{condition}[Mutual exclusivity conditions]{[B]MUTEX}
	Let $\psi_i \in \mathcal{P}$ be a countable sequence of states. Then the following are equivalent:
	\begin{itemize}
		\item $\psi_i$ are mutually exclusive
		\item $p(\psi_i | \psi_j) = \delta_{ij}$
		\item the entropy obeys $S(\sum_i p_i \psi_i) = - \sum_i p_i \log p_i$ for all $p_i$ such that $\sum_i p_i = 1$.
	\end{itemize}
\end{condition}
\begin{condition}[Conditional expectation from Born rule]{BR-BORN}
	The conditional probability $p : \mathcal{P} \times \mathcal{P} \to [0,1]$ is given by the Born rule. That is, $p(\psi|\phi) = \frac{\< \phi | \psi \>\< \psi | \phi \>}{\< \psi | \psi \>\< \phi | \phi \>}$.
\end{condition}
\begin{condition}[Von Neumann entropy]{BR-ENT}
	The entropy function $S : \mathcal{D} \to [0,+\infty]$ is given by the von Neumann entropy $S(\rho) = - \tr(\rho \log \rho)$.
\end{condition}
\begin{condition}[Orthogonality is mutual exclusivity]{BR-ORME}
	Orthogonal states are mutually exclusive.
\end{condition}
\begin{Theorem}
	Over \ref{QE-DENS}+\ref{[B]MUTEX}+\ref{[B]PMIX}, conditions \ref{BR-BORN}, \ref{BR-ENT} and \ref{BR-ORME} are equivalent. Moreover, \ref{BR-BORN} is independent of \ref{QE-DENS}+\ref{[B]MUTEX}+\ref{[B]PMIX}.
\end{Theorem}

This type of result is characteristic of Reverse Physics: structures normally packaged together in the standard formulation are separated and shown to represent different assumptions.

Condition \ref{[B]MUTEX} captures the idea that the same concept, the mutual exclusivity of preparations $A$ and $B$, can be characterized in two other equivalent ways. In terms of conditional probability, if we prepared $A$, there is zero probability of finding something that $B$ could have prepared; in terms of information entropy, we can use $A$ and $B$ to transfer a whole bit of entropy. Since these are necessary properties of mutual exclusivity, \ref{[B]MUTEX} is a base condition. The equivalence between \ref{BR-BORN} and \ref{BR-ENT} stems from the fact that the conditional probability between two states is an invertible function of the entropy of their equal mixture. Note that \ref{BR-BORN} implies that orthogonal states are mutually exclusive, and therefore \ref{BR-ORME}. Since every density operator can be orthogonally decomposed, \ref{BR-ORME} implies \ref{BR-ENT} through \ref{[B]MUTEX}.

The Born rule, then, has the implicit assumption that mutually exclusive preparations exist, which is not required by \ref{QE-DENS}. As a counterexample, consider the Bloch sphere and ball as the state and ensemble spaces for a two-state system. This state space assumes that the pure states are reachable. Now, suppose there is an entropy cutoff because of unavoidable noise. The reachable space, then, is a smaller ball centered at the origin. In that case, the state space $\mathcal{P}$ is the inner sphere, while $\mathcal{D}$ is the inner ball. While each state in $\mathcal{P}$ can still be reconstructed statistically through tomography, it cannot be confirmed with a single-shot measurement because of the noise. Similarly, no pair of preparations would allow the reliable transfer of one bit. Since the full and the cutoff balls are affinely isomorphic, \ref{QE-DENS} is still valid. In general, we do not see (yet) any fundamental reason why the existence of mutually exclusive states should be imposed a priori, and it may be that state spaces in a future more fundamental theory do not have this property.

\textbf{Probability requires mutual exclusivity.} The crucial insight is that the convex structure of the ensemble space is not enough to define probability; it requires this additional \textbf{entropic structure}. The mixture coefficients $p_i$ of a mixture $\sum_i p_i \psi_i$ can be interpreted as probabilities if and only if the $\psi_i$ are mutually exclusive. Kolmogorov probability, in fact, is based on the assumption that all the elements of the sample space are mutually exclusive \cite{BertsekasTsitsiklis2008}. In this light, contextuality is more than justified: there is no alternative.

The independence of the entropic structure from the convex structure of the ensemble space is present in classical mechanics as well: the Liouville measure is an additional structure on top of $\mathbb{R}^n$ and exactly corresponds to the entropy of uniform distributions in each region \cite{David2015,Ochs1976}. Therefore, it is not possible to reconstruct the entropy from the convex set of probability measures alone.

Theorems and derivations of the Born rule, like Gleason's theorem, or some approaches to Generalized Probabilistic Theories (GPT) often fail to make this distinction explicit \cite{Gleason1957,Plavala2023}. Many of these approaches, in fact, show that a unique affine linear functional with some features (e.g. zero and one antipodal ensembles) exists and identify it with probability. But the last move implicitly assumes mutual exclusivity, which may not be warranted. Moreover, classical mechanics requires the symplectic form in the definition of the state space, and since the Liouville measure cannot be reconstructed from the convex structure alone, these strategies cannot recover the classical state space. Similar, and in fact bigger, problems will exist for field theories.

\section{Unitary evolutions as deterministic and reversible processes}

We now turn our attention to time evolution, for which we have the following result \cite{Carcassi2026SchroedingerCharacterizations,Carcassi2026UnitaryEntropy}.
\begin{condition}[Evolution preserves statistical mixtures]{[B]EVMIX}
	The evolution of a statistical mixture is the statistical mixture of the evolutions. That is, let $\mathcal{D}$ be the ensemble space of the system. The evolution from each $t_0$ to each $t_1$, with $t_1\geq t_0$, is given by an affine map $\Phi_{t_0 \to t_1}:\mathcal{D}\to \mathcal{D}$ such that $\Phi_{t_0 \to t_0}=I$, $\Phi_{t_1 \to t_2}\circ \Phi_{t_0 \to t_1}=\Phi_{t_0 \to t_2}$ and $(\rho,t_0,t_1)\mapsto \Phi_{t_0 \to t_1}(\rho)$ is jointly continuous.
\end{condition}

\begin{condition}[Time-independent evolution]{EVTI}
	Evolution depends only on elapsed time. That is, $\Phi_{t_0 \to t_1}=\Phi_{t_1-t_0}$.
\end{condition}

\begin{condition}[Schrödinger equation]{DR-SCEQ}
	The evolution follows the equation $\frac{d}{dt} \psi(t) = \frac{H(t)}{\imath \hbar} \psi(t)$, where $H(t)$ is a self-adjoint operator.
\end{condition}

\begin{condition}[Unitary evolution]{DR-UNIT}
	Time evolution is represented by a strongly continuous one-parameter group of unitary operators. That is, $\Phi_t(\rho)=U_t\rho U_t^\dagger$, where $U_t:\mathcal{H}\to\mathcal{H}$, $U_0=I$, $U_{t+s}=U_tU_s$ and $U_t^\dagger U_t=U_tU_t^\dagger=I$.
\end{condition}

\begin{condition}[Bijective map]{DR-EV}
	Distinct past states map to distinct future states and vice-versa. That is, if $\psi_I$ and $\phi_I$ are two initial states, $\psi_F$ and $\phi_F$ are the corresponding evolved future states and $\psi_P$ and $\phi_P$ are the corresponding reconstructed past states, then $\psi_I = \phi_I \iff \psi_F = \phi_F \iff \psi_P = \phi_P$.
\end{condition}

\begin{condition}[Preservation of conditional probability]{DR-PROB}
	Conditional probability remains the same across past and future states. That is, if $\psi_I$ and $\phi_I$ are two initial states, $\psi_F$ and $\phi_F$ are the corresponding evolved future states and $\psi_P$ and $\phi_P$ are the corresponding reconstructed past states, then $p(\psi_I \mid \phi_I) = p(\psi_F \mid \phi_F) = p(\psi_P \mid \phi_P)$.
\end{condition}

\begin{condition}[Conservation of information entropy]{DR-INFO}
	Future and past ensembles have the same information entropy. That is, if $\rho_I$ is an initial ensemble, $\rho_F$ is the corresponding evolved future ensemble and $\rho_P$ is the corresponding reconstructed past ensemble, then $S(\rho_I) = S(\rho_F) = S(\rho_P)$.
\end{condition}

\begin{Theorem}
	Over \ref{QSTO:H}+\ref{[B]EVMIX}+\ref{EVTI}, conditions \ref{DR-SCEQ}, \ref{DR-UNIT}, \ref{DR-EV}, \ref{DR-PROB} and \ref{DR-INFO} are equivalent.
\end{Theorem}

This illustrates the role of explicitly stating the background assumptions: the conservation conditions become equivalent to unitary evolution only after evolution of mixtures and time independence have been assumed.

The base condition \ref{[B]EVMIX} captures the idea that time evolution preserves statistical mixtures. If the statistical mixture represents lack of knowledge of the initial conditions, then this lack of knowledge has to be transported to the final state. Condition \ref{EVTI} restricts to the case in which the process is time independent. In this case, if we have a bijection between pure states, the affine continuity will force the evolution to be a unitary group. In this sense, condition \ref{[B]EVMIX} does most of the work. Conditions \ref{DR-EV}, \ref{DR-PROB} and \ref{DR-INFO} just impose the bijection: \ref{DR-EV} does so directly; \ref{DR-PROB} through the Born rule, since $p(\psi|\phi) = 1 \iff \psi = \phi$; \ref{DR-INFO} through the entropy, because the entropy of an equal mixture of two pure states is an invertible function of the conditional probability between them.

The theorem can be extended to the time-dependent case by requiring the evolution to be differentiable instead of time independent (i.e. a differentiable unitary propagator). Physically, this effectively means that the time dependence is slow enough that, at some small scale, the evolution can be approximated by time-independent evolution.

\textbf{Applicability of unitary evolution.} The above results make it clear that the Schrödinger equation applies, and only applies, to deterministic and reversible systems: when the evolution of the system depends, and only depends, on the system itself. Any open evolution, then, is necessarily not unitary, and a measurement is necessarily not a closed evolution since the system must interact with the probe \cite{nielsen2010quantum,BreuerPetruccione2002}.

\textbf{Determinism and reversibility for ensembles.} If the state is an ensemble, deterministic and reversible evolution does not mean that there is no stochastic component in the system-environment interaction. The stochastic component can simply remain within the uncertainty of the initial ensemble, like for a stochastic equilibrium \cite{Nelson1966}. Therefore, the evolution can also be understood as the quasi-static evolution of stochastic equilibria \cite{Pavliotis2014,MaesNetocny2014}.

\section{Projections as equilibration processes}

\textbf{Measurements as equilibrations.} The following result characterizes nonselective projective measurements (i.e. the output of the measurement process before the readout) as equilibration processes \cite{Luders2006,Carcassi2026ProjectiveMeasurement}.

\begin{condition}[Nonselective projective measurement]{EQ-MEAS}
	The spectral equilibration process $\Phi_X$ associated with an observable $X=\sum_x xP_x$ returns the output of a nonselective projective measurement. That is, $\Phi_X(\rho)=\sum_x P_x \rho P_x$.
\end{condition}

\begin{condition}[Spectral equilibration]{EQ-LIND}
	The spectral equilibration process $\Phi_X$ associated with an observable $X=\sum_x xP_x$ can be characterized as the equilibration under a Lindblad equation with vanishing Hamiltonian and jump operator $X$. That is, $\frac{d\rho}{dt}=X\rho X-\frac{1}{2}\{X^2,\rho\}$ and $\Phi_X(\rho)=\lim\limits_{t\to\infty}\rho(t)$, where $\rho = \rho(0)$.
\end{condition}

\begin{condition}[Closest commuting ensemble]{EQ-HS}
	The spectral equilibration process $\Phi_X$ associated with an observable $X=\sum_x xP_x$ returns the ensemble commuting with $X$ that is closest to the incoming ensemble in Hilbert--Schmidt distance. That is, $\Phi_X(\rho)=\operatorname*{argmin}\limits_{\sigma\in\mathcal D(\mathcal H),\,[\sigma,X]=0}\|\rho-\sigma\|^2_{HS}$.
\end{condition}

\begin{condition}[Closest commuting ensemble in information]{EQ-KL}
	The spectral equilibration process $\Phi_X$ associated with an observable $X=\sum_x xP_x$ returns the ensemble commuting with $X$ that is closest to the incoming ensemble in relative entropy. That is, $\Phi_X(\rho)=\operatorname*{argmin}\limits_{\sigma\in\mathcal D(\mathcal H),\,[\sigma,X]=0}D(\rho\Vert\sigma)$.
\end{condition}

\begin{condition}[Entropy is finite]{[B]FINENT}
	All ensembles have a well-defined finite entropy.
\end{condition}

\begin{Theorem}
	Over \ref{QSTO:H}, conditions \ref{EQ-MEAS}, \ref{EQ-LIND} and \ref{EQ-HS} are equivalent. Over \ref{QSTO:H} constrained by \ref{[B]FINENT}, condition \ref{EQ-KL} is also equivalent to the others.
\end{Theorem}

A nonselective projective measurement removes the off-diagonal elements of the incoming state with respect to the spectral eigenspaces. A purely dissipative Lindblad process where $X$ is the jump operator does precisely that. Moreover, density operators that commute with $X$ are exactly those that are unchanged by nonselective projective measurements. Within those, the output is the closest one to the input in terms of the Hilbert--Schmidt distance, which is the one induced by the inner product between density matrices. One can alternatively characterize closeness using the relative entropy, if this is finite. Note that the relative entropy $D(\rho\Vert\Phi_X(\rho))$ can be infinite only if the entropy of the final state is infinite. Therefore, under the constraint \ref{[B]FINENT}, the conditions are equivalent. The equivalence, then, fails only because \ref{QSTO:H} fails to fully capture the physical requirements. This is why we still use the same prefix for \ref{EQ-KL}.

Note that, in the second part of the theorem, the background theory is not \ref{QSTO:H}+\ref{[B]FINENT}, the logical conjunction of the two conditions. This, in fact, would simply exclude the infinite case as the latter condition is not satisfied. The background theory, instead, is \ref{QSTO:H} constrained by \ref{[B]FINENT}, which means we allow the infinite-dimensional case and amend it by removing the ensembles that do not have finite entropy. Handling such restrictions exemplifies another type of investigation we need to pursue within Reverse Physics.

\textbf{Measurements as equilibration processes.} The first part of a measurement, the one responsible for the decoherence, can be understood as a purely dissipative equilibration process, meaning that all outputs are equilibria of the process and entropy increases if and only if the input is not already an equilibrium \cite{nielsen2010quantum,Preskill1998}. The difference from thermodynamic equilibration is in terms of the constraints satisfied through the process. The second part of a measurement, the discovery of the outcome and the update of the state, is identical to the classical case \cite{Ozawa1985}. In the classical case, we are always in a spectral equilibrium, so the spectral equilibration does nothing.

\textbf{Unitary evolution from measurements.} The above theorem shows that we can always understand a measurement in terms of an open system. The following shows that we can always understand unitary evolution as a sequence of infinitesimal projective measurements \cite{Carcassi2026UnitaryProjectiveEvolution}.

\begin{Definition}[State-adaptive projective process]
	A state-adaptive projective process is a process that acts as a projective measurement on each ensemble. That is, it is an affine map $\Psi: \mathcal{D} \to \mathcal{D}$ on the ensemble space such that, for every incoming ensemble $\rho$, there is a projective decomposition $\{P_a^\rho\}$ of the identity for which $\Psi(\rho)=\sum_a P_a^\rho\rho P_a^\rho$. The projective decomposition may depend on the incoming ensemble.
\end{Definition}

\begin{condition}[Projective evolution]{DR-PROJ}
	Time evolution is the limit of repeated applications of an infinitesimal state-adaptive projective process that allows perfect prediction and perfect reconstruction. That is, evolution over elapsed time $t$ is obtained by applying the same process of duration $t/n$, $n$ times, in the limit $n\to\infty$. Moreover, the probabilities of the predicted and reconstructed state sequences tend to one.
\end{condition}

\begin{Theorem}
	Over \ref{QSTO:H}, conditions \ref{DR-UNIT} and \ref{DR-PROJ} are equivalent.
\end{Theorem}

A state-adaptive projective process is effectively a measurement process where the choice of measurement depends on the initial state. Each incoming state, even a mixed state, sees a projective measurement, but each state may see a different projective measurement. To recover the unitary evolution, one repeats the process at each infinitesimal time step in a way that makes the change infinitesimal. This is similar to a quantum Zeno effect or quantum Zeno transport \cite{FacchiPascazio2008,BurgarthEtAl2013}.

As a simple way to understand this, suppose one starts with an eigenstate $\psi$ of some observable $X$. Both the state and the observable will evolve so that $\psi(t) = U_t \psi$ will remain an eigenstate of $X(t) = U_t X U^\dagger_t$. In this sense, the unitary process is ``measuring'' the evolved observable at every instant. The above result formalizes this intuition.

\textbf{Forward and backward measurement problem.} The conclusion is that we can understand the measurement process as an equilibration of an open quantum system, which is a special case of unitary evolution for the system plus the environment; but we can also understand unitary evolution as a quasi-static process made of infinitesimal equilibrations at each time step.\footnote{The quasi-static interpretation of unitary evolution is consistent with $S$-matrix calculations in field theory, where scattering processes that may last much less than a femtosecond are modeled with instates at minus infinity and outstates at plus infinity \cite{Weinberg1995}.} That is, this equivalence addresses both the forward (i.e. unitary $\to$ projection measurements) and backward (i.e. projection measurement $\to$ unitary) problems. The key is, again, assuming that states represent statistical ensembles, as only in this case can one define equilibration processes.

\section{Classical mechanics as high-entropy limit}

The following result establishes the connection between classical and quantum mechanics.

\begin{condition}[One DOF]{1DOF}
	The system is fully characterized by conjugate position and momentum, together with their Lie bracket (i.e. $\{q,p\}=1$ in classical mechanics and $\frac{[X,P]}{\imath\hbar}=I$ in quantum mechanics).
\end{condition}
\begin{condition}[Small $\hbar$]{CL-HBAR}
	The value of the constant $\hbar$ can be considered small for the problems at hand.
\end{condition}
\begin{condition}[High entropy]{CL-HENT}
	The entropy can be considered high for the problems at hand.
\end{condition}
\begin{condition}[Classical limit]{CL-CST}
	The system can be approximated by a classical system for the problems at hand.
\end{condition}
\begin{Theorem}
	Over \ref{QSTO:H}+\ref{1DOF}, the conditions \ref{CL-HBAR}, \ref{CL-HENT} and \ref{CL-CST} are equivalent.
\end{Theorem}

This is an example of cross-theory comparison that makes explicit the shared physical structure being preserved: the Lie bracket.

Since \ref{1DOF} is part of the background theory, it must hold in both the quantum and classical descriptions. Therefore, in the classical limit, the quantum commutator must reduce to the classical Poisson bracket. In the Weyl--Wigner representation, the quantum commutator is represented by the Moyal bracket \cite{Moyal1949}. A result from deformation theory states that, up to a change of representation, the Moyal bracket is the unique non-trivial formal differentiable deformation of the Poisson bracket on canonical phase space \cite{BayenEtAl1978I,Gutt1979,OvsienkoRoger1992}. Moreover, it reduces to the Poisson bracket when $\hbar$ becomes negligible \cite{hillery1984distribution}. Within this setting, conditions \ref{CL-CST} and \ref{CL-HBAR} therefore describe the same limit.

To connect this limit to entropy, consider an open quantum system with the Lindblad evolution given by $H=0$ and $L=a^\dagger$ \cite{aop-phys-ClassicalHighEntropyLimit}. After some time $t$, this channel stretches both position and momentum by $\sqrt{\lambda} = e^{\frac{\gamma}{2}t}$, where $\gamma$ is the Lindblad rate coefficient, and it increases entropy and suppresses the negative regions of the Wigner function in the limit $\lambda\to\infty$. The same transformation can instead be described using the rescaled observables $\hat X=X/\sqrt{\lambda}$ and $\hat P=P/\sqrt{\lambda}$. These satisfy $[\hat X,\hat P]=\imath\hbar/\lambda$, so increasing the entropy through $\lambda\to\infty$ is equivalent to taking the effective value of $\hbar$ to zero. In entropic terms, this shifts the entropy assigned to the pure-state scale from zero to $-\log\lambda$, which tends to minus infinity. Every finite entropy is therefore high relative to the entropy of pure states.\footnote{This is analogous to setting $c\to\infty$ in the non-relativistic limit: one does not change the physical constant, but expresses the approximation by moving the limiting scale infinitely far away.} Conditions \ref{CL-HBAR} and \ref{CL-HENT} therefore describe the same limit.

\textbf{Reinterpret Dirac's correspondence principle.} The standard way to quantize a classical theory is through Dirac's correspondence principle, which formally substitutes commutators for Poisson brackets \cite{Dirac1925}. The above result makes the physics behind this correspondence clear: we are looking for a theory that has a lower bound on the entropy (i.e. zero on pure states) but recovers classical mechanics at high entropy.\footnote{This recovery assumption is key, as one can go to high entropy in a way that distorts the Lie algebra.}

\textbf{Fixing thermodynamics.} Historically, many failures of classical theories came from thermodynamics and statistical mechanics \cite{Taschetto2025,EisbergResnick1985,FayngoldFayngold2013}. Given that classical theory allows negative and arbitrarily low entropy, which are problematic for thermodynamics, this should not be surprising \cite{Wehrl1991}. The lower bound on the entropy imposed through quantization, then, is needed to fix those problems. What is remarkable is how many other results follow.

\section{States as ensembles in equilibrium}

This result shows that quantum states can be characterized as three types of equilibria, and that the same mathematical condition can have multiple physical consequences.

\begin{condition}[Dynamic equilibrium]{EE-DYN}
	Every ensemble $\rho$ is the dynamic equilibrium of some unitary process.
\end{condition}

\begin{condition}[Spectral equilibrium]{EE-MEAS}
	Every ensemble $\rho$ is the output of some spectral equilibration process (i.e. nonselective projective measurement).
\end{condition}

\begin{condition}[Thermodynamic equilibrium]{EE-THER}
	Every full-rank ensemble $\rho$ is the thermodynamic equilibrium for some Hamiltonian. Every non-full-rank ensemble is an approximate thermodynamic equilibrium for some Hamiltonian.
\end{condition}

\begin{Theorem}
	Condition \ref{QSTO:H} implies \ref{EE-DYN}, \ref{EE-MEAS} and \ref{EE-THER}.
\end{Theorem}

For \ref{EE-DYN}, set $H=\rho$. Then $[H,\rho] = 0$, so $\rho$ is stationary. For \ref{EE-MEAS}, choose $X = \rho$. Then $[X,\rho] = 0$, so $\rho$ is a fixed point. For \ref{EE-THER}, if $\rho$ is full rank, choose $H=-\log\rho/\beta$ \cite{Wilming2017}. Then $e^{-\beta H}=\rho$ and $Z=\tr(\rho)=1$. Otherwise, use the same definition on the support of $\rho$ and assign energies $kn^2$ to an orthonormal basis of its kernel. The corresponding normalized Gibbs states converge to $\rho$ in the Hilbert--Schmidt norm as $k\to\infty$, since $\sum_n e^{-\beta kn^2}\to0$.

\textbf{States as ensembles.} Since the above result holds in quantum theory, it has to hold in any interpretation. Therefore, an ensemble interpretation is not just minimal \cite{Ballentine1970,HomeWhitaker1992}, but it is unavoidable in the sense that it is already baked into the mathematical structure. The math and physics, then, point to the fundamentality of the concept of ensemble which, in retrospect, is philosophically obvious. If scientific laws are about relationships that can always be experimentally reproduced, the objects of those relationships are not the single instances of the replications, but rather the infinite collection of all past, present and future replications. The state of the system in a physical law, then, cannot describe a complete picture of the system, but only those aspects that can be replicated experimentally. The zero bound on the entropy precisely divides what information is reliably accessible through all experimentation. This lower bound is not epistemic, but determined by the physical processes that act on the system under the chosen conditions. We aim to fully develop these philosophical ideas in the future, hopefully in collaboration with other philosophers of science, and to use them as a foundation for a more cohesive mathematical framework.

\section{Conclusions}

In this paper we have shown how Reverse Physics leads to a series of results on the foundations of quantum mechanics that help better characterize each element of the theory. It does so by breaking up the theory into a series of conditions, some mathematical and some physical, and studying their logical relationships. It can be used to distinguish base conditions, those that must be present in every well-defined physical theory, from those that represent additional, more restrictive assumptions. It allows one to study logical equivalence between mathematical and physical conditions, highlighting when the mathematical framework fails to be physically significant. Most importantly, all the results stand on their own as self-contained theorems, and are formulated within the standard vector-space representation of quantum mechanics, making them accessible to the widest possible audience. Moreover, similar results are available for classical mechanics, allowing cross-theory comparisons.

This means that the results can be used in many different ways for many different goals. The teachers or students who simply want to have a better understanding of the framework, the philosophers who want to develop their interpretation and extend the theory with a well-defined ontology, the researchers who want to reconstruct quantum theory from different premises, the physicists who want to develop a new theory: all of them can benefit from a more detailed understanding of the relationships among the many different conditions studied. As the methodology matures, we hope that many other results can be incorporated with the same style, making it much easier to share fundamental insights and results.

\authorcontributions{Conceptualization, methodology, G.C.; validation, C.A.A. and T.T.; writing---original draft preparation, G.C.; writing---review and editing, C.A.A. and T.T.; funding acquisition, C.A.A. and G.C. All authors have read and agreed to the published version of the manuscript.}


\funding{This paper was made possible in part through the support of grant \#62847 from the John Templeton Foundation.}


\institutionalreview{Not applicable.}

\informedconsent{Not applicable.}

\dataavailability{This study generated no data.} 

\acknowledgments{This paper is part of the ongoing open research program \emph{Assumptions of Physics} \cite{aop-book}. The authors wish to thank Andrea Oldofredi for helpful discussions.  During the preparation of this manuscript, the authors used ChatGPT 5.6 Sol for the purposes of proof verification, literature search and consistency checks. The authors have reviewed and edited the output and take full responsibility for the content of this publication.}

\conflictsofinterest{The authors declare no conflicts of interest.} 




\isPreprints{}{
\begin{adjustwidth}{-\extralength}{0cm}
} 

\reftitle{References}




\bibliography{bibliography,assumptionsofphysics,MoyalBracket,Introduction,ReversePhysics}

%


\isPreprints{}{
\end{adjustwidth}
} 
\end{document}